# Fog Data: Enhancing Telehealth Big Data Through Fog Computing

Harishchandra Dubey[#], Jing Yang, Nick Constant, Amir Mohammad Amiri, Qing Yang[+], Kunal Makodiya[*]
Department of Electrical, Computer, and Biomedical Engineering
University of Rhode Island, Kingston, RI 02881, USA
[#]dubey@ele.uri.edu, [+]qyang@ele.uri.edu, [*]kunalm@uri.edu

## ABSTRACT
The size of multi-modal, heterogeneous data collected through various sensors is growing exponentially. It demands intelligent data reduction, data mining and analytics at edge devices. Data compression can reduce the network bandwidth and transmission power consumed by edge devices. This paper proposes, validates and evaluates *Fog Data*, a service-oriented architecture for Fog computing. The center piece of the proposed architecture is a low power embedded computer that carries out data mining and data analytics on raw data collected from various wearable sensors used for telehealth applications. The embedded computer collects the sensed data as time series, analyzes it, and finds similar patterns present. Patterns are stored, and unique patterns are transmited. Also, the embedded computer extracts clinically relevant information that is sent to the cloud. A working prototype of the proposed architecture was built and used to carry out case studies on telehealth big data applications. Specifically, our case studies used the data from the sensors worn by patients with either speech motor disorders or cardiovascular problems. We implemented and evaluated both generic and application specific data mining techniques to show orders of magnitude data reduction and hence transmission power savings. Quantitative evaluations were conducted for comparing various data mining techniques and standard data compression techniques. The obtained results showed substantial improvement in system efficiency using the *Fog Data* architecture.

## 1. INTRODUCTION
With the increasing use of wearable sensors in healthcare and biomedical applications, we are living in the data-driven world. As a result, we are presented with many challenges in dealing with big data. "Big Data" is characterized by high-volume, high-variety, and high-velocity information that demands efficient and innovative processing for enhanced insight and decision-making [1]. For example, in the healthcare domain, telehealth enables the use of sensors within or on the human body. Wearable sensors such as ECG and activity monitors are a specific type of medical sensors placed on the human body allowing non-invasive, unobtrusive, 24/7 data collection for health monitoring. The increasing number of ambient devices installed in homes that surround the human body provide telehealth interventions to citizens seeking affordable healthcare while remaining in touch with medical practitioners remotely. Such telehealth application is a typical example of big data application that collects a large volume of data with a variety of information requiring real time and fast processing to provide the best healthcare. There are several challenges in deploying telehealth applications. Firstly, designing information sensing nodes in body sensor networks (BSNs) is a challenge that is eased by using wearable sensors, e.g., smartwatches. The second challenge is a collection, storage and analysis of large amount of heterogeneous, multi-modal, distributed, and scalable data sets, known as big medical data. It is inefficient and impractical in several applications to use traditional architectures and algorithms for storing and processing such data. The third challenge is energy efficiency of wearable and portable edge devices used in a telehealth solution. Since BSNs are powered by batteries, to provide uninterrupted monitoring of patients, these batteries should not be frequently recharged. Therefore, low power consumption is critical for BSNs. Typically, data storage and data transmission consume a significant amount of energy suggesting the benefits of quick and preliminary data analytics to reduce the amount of necessary data to be stored and transmitted.

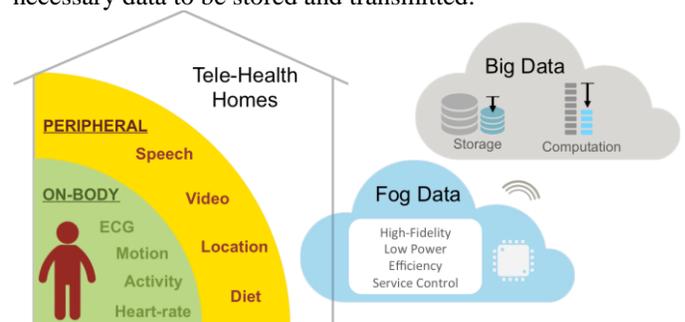

**Figure 1. Fog Data, a service-oriented architecture to reduce storage requirements and to increase the overall efficiency of telehealth big data solutions.**

The authors are grateful to the anonymous reviewers for providing comments and suggestions that improved the quality of the paper. This research is supported in part by NSF grants CCF-1439011 and CCF-1421823. Any opinions, findings, and conclusions or recommendations expressed in this material are those of the author(s) and do not necessarily reflect the views of the National Science Foundation.

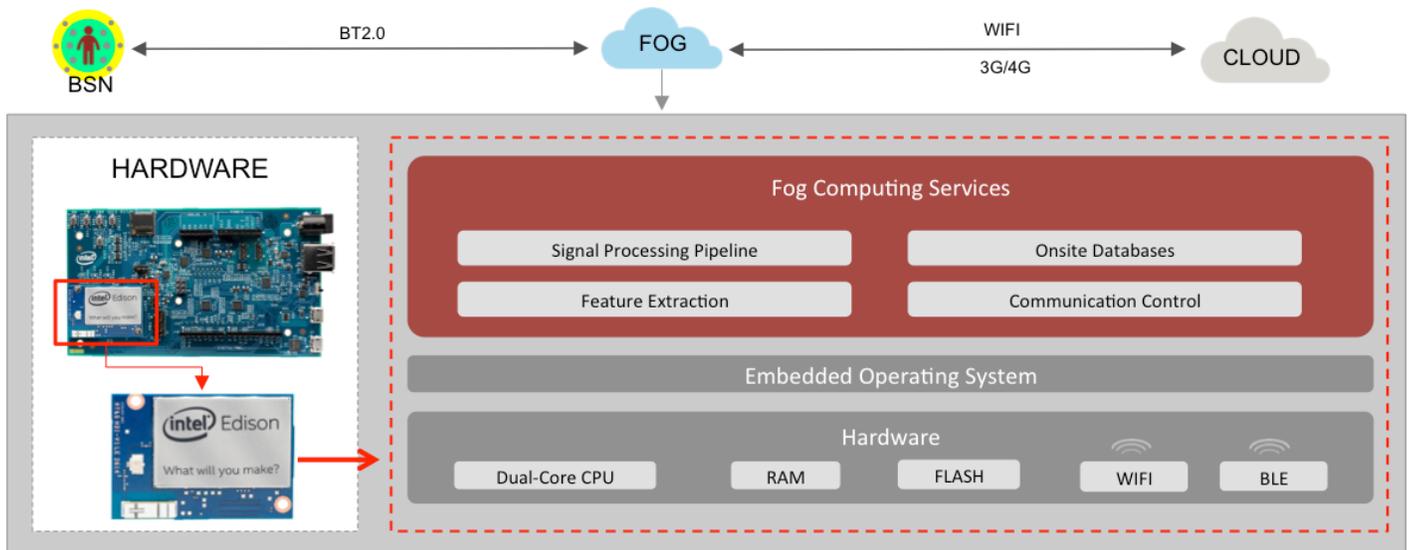

**Figure 2. A system architecture of Fog Data.**

*Fog Data*, a new system architecture based on Fog computing concept, is presented in this paper as a means to tackle these challenges. The proposed architecture is a service-oriented Fog computing architecture interfacing in-home telehealth devices facilitating the collection, storage, and analysis of large amount of heterogeneous, multimodal, distributed and scalable data sets for person-centered health monitoring. The unique feature of *Fog Data* architecture is its ability to carry out onsite data analytics to reduce the amount of data to be stored and transmitted to the cloud. As shown in Figure 1, wearable sensors and ambient devices at home collect the necessary information as raw data. The raw data is generally in the form of time series signals that are transmitted to energy efficient embedded computer, referred to as Fog computer. The Fog computer handles transmitting necessary data to the cloud after preliminary analysis and filtering.

To validate the proposed architecture, we have designed and implemented a system prototype based on Intel® Edison embedded processor connected to the wearable telehealth systems. Using the working prototype, we have carried out extensive experiments using real data sets collected from patients and a set of open source data. Experimental results have shown an order of magnitude in data reduction. This paper made the following contributions:

a) A new Fog computing architecture was proposed emphasizing on data reduction, low power consumption, and high efficiency;
b) A working prototype was built using Intel®'s Edison embedded processor to demonstrate the efficiency of the new architecture;
c) Extensive experiments were carried out to show that the new system substantially reduced the amount of data that needs to be stored and transferred;
d) Case studies were performed on telehealth applications namely speech disorders and ECG monitoring.

## 2. RELATED WORKS

### 2.1. TeleHealth and Medical Big Data

An increasing population with chronic diseases and old age along with the rise in medical costs has created a demand to extend the healthcare services from hospital to home with a focus on efficiently managing disease and overall wellbeing of patients. In the last decade, this scenario has given rise to a new paradigm called "telehealth" that allows patient health monitoring and disease management in non-clinical settings such as private homes, nursing homes, and assisted living. Telehealth infrastructure consists of Body Sensor Network (BSN) combining wearable sensors and personal area network [2]. To further illustrate telehealth here are some of the notable wearable telehealth systems: 1) VitalConnect is a band-aid style healthpatch to collect continuous vital signs in remote settings [3]. 2) Philips has launched an adhesive patch to monitor Chronic Obstructive Pulmonary Disease (COPD) [4]. 3) Fitbit, a wristworn sensor contains a motion sensor for estimating health indices such as step counting, calories, and sleep quality [5].

Due to sensor-rich infrastructure, telehealth generates medical big data for remote diagnosis and clinical interventions. For example, EchoWear [6] that uses a modern smartwatch to provide a home-based speech tele-therapy that generates approximately 100Mb data per day per patient. This is an example of wearable single-modality data. The data complexity increases tremendously for multi-modal data [3]. Hence, gaining knowledge from the data becomes tedious and cumbersome. Wearable sensor data contains valuable information. However, they also carry non-deterministic errors such as motion artifacts, data corruption issues, and unwanted signals that are also uploaded increasing storage requirements and power consumption significantly. The model of sending raw medical data to the cloud for big data analysis is becoming inefficient, time consuming, and expensive. In the following subsection, we describe how the Fog computing could play an important role to increase the efficiency and reduce storage requirements for medical big data solutions.

## 2.2 Fog Computing as a Smart Gateway

Fog computing is a new paradigm of computing research coined by Cisco Inc. in 2012 [8]. Conventionally, the Fog is described as a cloud close to the earth's ground plane. The Fog computing means deploying computing services on remote devices to facilitate low latency, high efficiency, and high reliability for end-user applications spanning from healthcare to automobile to smart cities [9]. Fog computing is a smart gateway that handles performing various tasks:

- Local Connectivity: Fog gateway directly connects with wearable sensors to acquire data and to provide actuations such as alerts and notifications.
- Computation: Fog gateway processes the incoming data to produce medical-grade logs that are sent to the cloud for comprehensive analysis. The computation varies from simple filtering to complex wavelet analysis.
- Onsite Database: It creates a local database containing features and clinical parameters that can be queried internally and externally.
- Data Security: It allows users to deploy security layer to protect the data and user identity.

As it can be seen from the possible operations, Fog computing is a promising approach in the context of big telehealth data. In this paper, we discuss our implementation of Fog computing infrastructure and demonstrate two case studies showing the performance of Fog Computing to tackle wearable telehealth data.

## 3. THE ARCHITECTURE OF FOG DATA

Figure 2 shows the architecture of *Fog Data* as a cascade of three sub-systems, namely a body sensor network (BSN) for data acquisition, a Fog gateway computer for onsite processing, and the cloud server for storage and back-end analysis. The Fog computing services connects the patients with wearable sensors to a physician who diagnose, evaluate, and treat the patient. The Fog computer provides a smart gateway facilitating the reduction in storage space in the cloud and Internet bandwidth needed to send the clinically relevant information to the cloud. The Fog computer is connected to the Internet that allows on-demand real-time control of commands needed for computation of relevant features from the health data accumulated at the Fog computer. The *Fog Data* is a generic architecture that allows a wide variety of wearable sensors such as smartwatch, wearable ECG system, and pulse glasses to be used for acquisition of health data. The wearable sensors possess limited memory and computing resources. Consequently, these cannot accumulate the data acquired in real-time. For accumulating the data resulting from continuous monitoring of patients, *Fog Data* architecture has Fog Computing Services providing flexible software routine that perform on-demand, real-time accumulation of data, processing of data for extracting clinically relevant features or mining pattern in acquired data. Depending on the requirements of the healthcare application, the data collected from wearable sensors or the extracted features or the index of the pattern in the time-series are sent to the cloud. The cloud is the final sub-system in the chain that stores all data or features from Fog Computer. The Fog Computer has limited storage and computing resources as compared to computing resources available in the cloud. Thus, the algorithm and methods applied to the acquired data in Fog Computer need to be computationally simple and yet clinically relevant. This two stringent requirement narrows down the signal processing and machine learning algorithms that can be realized on Fog Computer. The cloud executes complex algorithms that are important for clinical diagnosis, monitoring, and treatment. The data is deleted from Fog Computer once it is sent or the extracted features are sent to the cloud. The cloud securely stores the data along with appropriate log files. Continuous monitoring of patients results in huge amount of data. The cloud back-end can analyze the received data by scalable speech processing and scalable machine learning methodologies where similar algorithms process different chunks of healthcare big data for inference and diagnosis regarding patients current state of health and any improvement in health over the time.

### 3.1 Information Flow in Fog Data Architecture

The data accumulated in *Fog Data* is time-series along with time-stamps from wearable sensors. Thus, it is known what data is collected and at what time. The time-series healthcare data accumulated at *Fog Data* can be speech or ECG data or other clinical vital signs. The Fog Computer Services processes the raw data collected from health monitoring and converts it into features/patterns along with time-stamps.

*1) Log files*

The data collection is accompanied with a log file that stores the time-stamps of data acquisition along with other important factors such as the percentage of battery power remaining at that time. The log files are .xls files that can be accessed by the physicians, patients and the engineers developing the algorithm for the wearable healthcare systems based on *Fog Data* architecture. These log files help in debugging the software programs for certain use cases.

*2) Sensing*

Using wearable sensors such as smartwatch creates a pleasant interface for a continuous collection of speech and other healthcare data that can be utilized for health monitoring and diagnosis by expert physicians and clinicians. Similarly, smart clothing based ECG sensors acquire the ECG signals containing clinically important information about the heart diseases.

*3) Methods for acquiring data*

We have compared several plausible ways to collect data from multi-modalities. The data acquired by BSN that employs wearable sensors interacting with nearby-placed portable data aggregators such as tablet or smartphone. Using such devices in data collection eliminates the need to have custom hardware devices. The EchoWear is a software framework consisting of applications running on smartwatch and tablet/phone developed and validated by authors in [6]. We have used EchoWear framework for acquiring speech data in case study 1 discussed in Section 6.

## 4. METHODS AND ALGORITHMS

This section explains the methods used for analysis of healthcare data on Fog Computer (Intel® Edison). These methods were implemented as software routines running on

Intel® Edison. The chosen methods achieve the data analytics suited for case studies developed in Section 6.

## 4.1 Dynamic Time Warping

Dynamic time warping (DTW) is an algorithm for mining patterns in time-series data. It has been used for various applications such as business, finance, single word recognition, walking pattern detection, and analysis of ECG signals. Usually, the Euclidean distance is used to measure the distance between two points. Euclidean distance fails to detect similarity between similar and out-of-phase series. On the other hand, DTW can detect similarity between two series regardless of different length, and phase difference. It builds an adjacency matrix then finds the shortest path across it [13, 23, 24].

## 4.2 Clinical Speech Processing Chain (CLIP)

Clinical speech processing chain (CLIP) is a series of filtering operation applied to speech data for computing the clinically relevant metric such as loudness and fundamental frequency.

*1) Loudness*

The perceived loudness level of a speech sound handles intensity sensation. Authors in [6], [17] used the Zwicker's method for loudness computation valid for time varying sound that is standardized as DIN 45631/A1 (2008). The loudness is a mathematical quantity that depends on amplitude, frequency-content, and time-duration of the signal. The loudness is most sensitive to changes in amplitude, followed by frequency-content and least dependent on time-duration of sound. The Fog processor has limited computing resources. Consequently, using Zwicker's model for computing loudness was not advisable as a first step. We instead used a coarse estimate of loudness based only on the amplitude of the speech signal. Loudness is approximated as the root mean squared power of speech signal in our experiments since the loudness has a strong dependence on signal amplitude.

*2) Fundamental Frequency*

In *Fog Data* architecture, the Fog processor Intel® Edison executes the algorithm for pitch (fundamental frequency) estimation after receiving the speech signal. In this paper, we used average magnitude difference function (AMDF) based method for pitch estimation [17]. The AMDF is similar to the autocorrelation function, but it does not need multiplication, unlike the autocorrelation function. Pitch detection based on AMDF has low computational cost facilitating simple implementation on Fog processor (Intel® Edison). We do the low-pass filtering of speech before AMDF processing for reducing the background noise embedded in the speech signal. AMDF based pitch estimation is slightly inaccurate as compared to SWIPE and Pratt's method for the fundamental frequency used in [6] and [17]. Since fundamental frequency doubling or halving errors are common for AMDF, we use median filtering to refine the pitch estimates based on AMDF.

## 4.3 Compression

We have used GNU zip as compression and decompression algorithm in our experiments [27]. The compression of data is done on the Fog processor that sends the compressed data to the cloud. The cloud back-end receives the compressed data and decompresses it before processing. GNU zip is well-known compression and decompression program originally developed for GNU Project in 1992. It is used in UNIX and Linux environment. The compression and decompression programs such as bzip2 and 7-zip provide higher compression ratio. However, these programs need more CPU power and RAM resource that is not suitable for embedded Fog processors [28]-[29].

## 4.4 Data and Bandwidth Reduction

The data transmission from Fog to the cloud and memory required for storage in Fog computer and the cloud are the crucial benefits offered by *Fog Data* architecture. Fog processor reduces data transmission and storage by local processing the healthcare data. For instance, clinical speech processing, Pan-Tompkins algorithm for ECG and DTW based pattern matching algorithms are implemented in Fog processor (Intel® Edison). However, the idea of local processing is invariant on the type of healthcare data and methods needed for processing. The results discussed in this paper are obtained from methods executed on Fog computer. DTW was programmed in C. CLIP has python front-end for C back-end. The methods can be used for another processor similar to Intel® Edison.

## 4.5 Power Consumption

Power consumption is an important component in Fog computing especially for portable battery-powered sub-systems. *Fog Data* is designed to be low-power architecture. We measure the power consumption of Intel® Edison that is 862 mW for DTW computations, and 937 mW for encrypted data transmission through Wi-Fi. The power consumption of *Fog Data* architecture is much lower than Raspberry Pi. The DTW computations were done occasionally and for a short time. Consequently, a standard 9V battery can supply sufficient power for at least 4 hours. Also, data reduction has reduced the power needed for data transmission (from Fog to the cloud) and storage power in the cloud.

## 5. EXPERIMENTAL SETUP

We used an Intel® Edison as Fog computing development platform in our experiments [18]. It is a low-power device powered by rechargeable lithium-ion battery or plug-in AC adapter [19]. The Intel® Edison is a System on Chip (SoC) that includes a 22nm dual-core, dual-threaded 500MHz Intel® Atom™ CPU and a 100MHz Intel® Quark™ microcontroller. There is 1GB LPDDR3 memory, and 4GB eMMC flash storage on board. It can connect to the Internet through WiFi interface that supports IEEE 802.11 a/b/g/n standards. The board consumes 35 mW while idle with WiFi open. To run our programs on Intel® Edison, we installed ubilinux [20]. The Python interpreter version 2.7.2 was installed. Intel® Edison can also communicate to tablet or smartphone through Bluetooth. Low-level C programs packed with Python wrappers for execution on Fog processor. The DTW algorithm was implemented in C program based on open source projects UCR Suite and libdtw [21]-[22]. The UCR Suite DTW implementation can search and analyze time series signals at real-time processing speed. There are methods for accelerating DTW processing that require special hardware support. The UCR Suite is one of the fastest software implementations. In our experiments, the patterns to be searched are decided by clinical requirements, e.g., in speech disorder case study described in section 6, 40ms speech segment is chosen as the reference pattern. For further reduction in network bandwidth requirements and improved accuracy, we use a waiting window in DTW algorithm. In case multiple similarity patterns were detected in the window, we send only the most similar one to the back-end server.in DTW algorithm. In case multiple similarity patterns were detected in the window, we send only the most similar one to the back-end server.

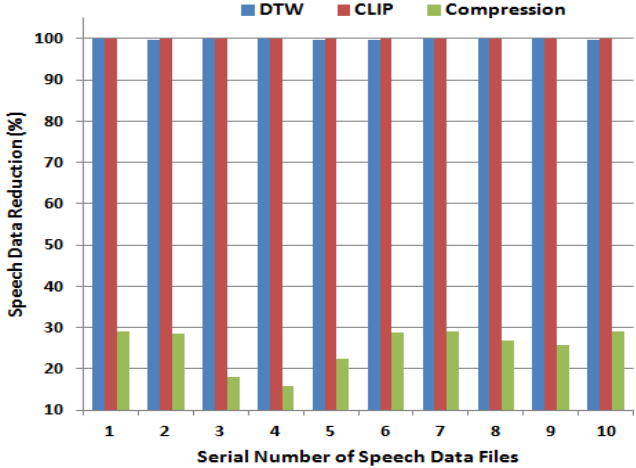

Figure 3. The percentage data reduction achieved by using dynamic time warping (DTW), clinical speech processing (CLIP) and GNU zip compression on 10 speech files collected from Parkinson's disease patient by smartwatch-based system developed in [6].

## 6. FOG DATA: CASE STUDIES

The first case study is for speech monitoring of patient with Parkinson's disease (PD). PD patients perform speech exercises in home settings. The Android smartwatch is used for acquiring speech signal and transferring it to a Fog computer (Intel® Edison) where speech is processed. In second case study, the average loudness and average fundamental frequency. The DTW performs close to CLIP with more than 99% data reduction. Both CLIP and DTW are lossy processes so they cannot be reversed back to give the speech files. The GNU zip is lossless compression and it can be reversed to generate the original speech time-series in the cloud. This reversibility comes with lower data reduction rate compared to CLIP and DTW. Fog computer access the MIT-BIH database through internet and applies Pan-Tompkins algorithm for QRS detection in ECG signals. Finally, the processed data is uploaded to the cloud for long-term storage and further analysis in both cases.

### 6.1 Case Study 1: Telehealth of Speech Motor Disorders

Approximately 7.5 million people in USA have speech disorders resulting from a variety of diagnoses. Speech-language pathologists (SLPs) are involved in the evaluation, diagnosis, and treatment of people with dysarthria [10], [12]. The traditional voice therapy is delivered in clinic by SLPs. Remote delivery of speech treatment is challenging due to inherent limitations. The EchoWear system facilitates the remote speech treatment [6]. EchoWear extracts the speech features such as loudness and pitch that can be accessed by SLPs for customizing the speech exercises as per the patient's needs. The speech data from smartwatch is sent to Fog gateway in *Fog Data* architecture. The Fog computer Intel® Edison processes the speech signals by methods described in Section 4. The extracted features and pattern indices are sent to the cloud by Fog computer. Figure 3 shows the data reduction achieved by feature extraction, pattern mining and compression. CLIP achieves highest amount of data reduction as it converts speech time-series into two features namely average loudness and average fundamental frequency. The DTW performs close to CLIP with more than 99% data reduction. Both CLIP and DTW are lossy process so they cannot be reversed back to give the speech files. The GNU zip is lossless compression and it can be reversed to generate the original speech time-series in the cloud. This reversibility comes with lower data reduction rate compared to CLIP and DTW.

### 6.2 Case Study 2: Telehealth of ECG Monitoring

Heart diseases are one of the major chronic illness making dramatic impact on productivity of affected individuals. The electrocardiogram (ECG or EKG) is a diagnostic tool to assess the electrical and muscular functions of the heart. The ECG signal consists of components such as P wave, PR interval, RR interval, QRS complex, pulse train, ST segment, T wave, QT interval and infrequent presence of U wave. The P wave, T wave and QRS complex are searched using DTW for a large number of ECG data sets. The last section of this case study will discuss the data reduction using DTW and GNU zip compression on ECG data. The goal of our experiment is to detect arrhythmic ECG beats or QRS changes using QRS complex and the RR interval measurements. The ECG data is fed to the Fog computer from Internet-based database. The Fog computer extracts QRS complex from ECG signals using real-time signal processing implemented in Python on Intel® Edison. The Pan Tompkins algorithm is used for detection of QRS complex [26]. Figure 4 illustrates the QRS detection using Pan-Tomkins algorithm on Intel® Edison using MIT-BIH Arrhythmia data [30]. The ECG signal containing 2160 samples take 1 second of processing time on Intel® Edison Fog computer. Thus, *Fog Data* architecture is well suited for real-time ECG monitoring. The DTW indices showing the location of these patterns in ECG time-series is sent to the cloud. Further, we used GNU zip program to compress the original ECG time series. The compressed ECG data files are then sent to the cloud. Figure 5a shows the data reduction resulting from DTW based pattern mining with compression. Similar to speech data, DTW reduces ECG data by more than 98% in most of the cases while compression reduces around 91%. We can see the difference in reduction efficiency is smaller than that in speech data shown in Figure 3. Figure 5b shows the execution time (in seconds) for Pan-Tompkins based QRS detection implemented in Python on Intel® Edison Fog computer. The size of the data sets ranges from 16.24 kB to 36.45 kB. The execution time increases with an increase in file size. The time taken is always less than 15 seconds. This validates the efficacy of *Fog Data* architecture for real-time ECG monitoring.

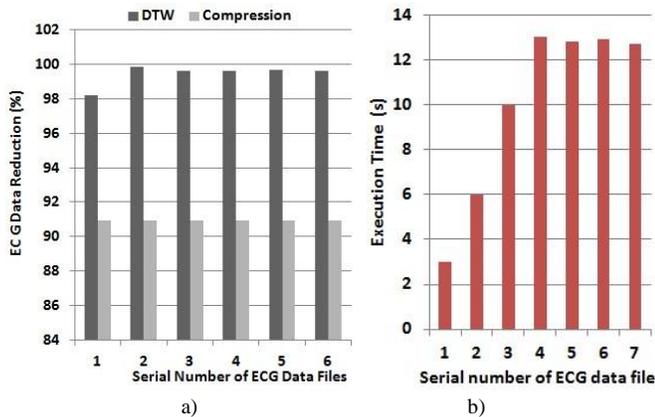

Figure 5. Results for ECG data obtained from MIT-BIH Arrhythmia Database. a) Comparison of data reduction resulting from DTW and GNU zip compression b) Execution time (in seconds) for Pan-Tompkins based QRS detection on Intel Edison Fog computer.

## 7. CONCLUSIONS

We have validated the use of *Fog Data* architecture for two important healthcare problems namely speech disorders and ECG. The solutions based on *Fog Data* architecture has the potential to reduce logistics requirements for telehealth applications in addition to the reduction in required cloud storage and transmission power at edge devices. There are several aspects that can be investigated in future. Speech features like shimmer, jitter and sensory pleasantness that are useful for quantification of perceptual speech quality can be computed in addition to loudness and fundamental frequency for speech disorders. It would be interesting to use the Fog Data architecture for in-home validation studies of EchoWear developed in [6].